\documentclass[aps,10pt,twocolumn,superscriptaddress, nofootinbib]{revtex4-2}
\usepackage{mathrsfs, amssymb, amsmath}  
\usepackage{epsfig, cancel}
\usepackage{footmisc}
\usepackage{latexsym}
\usepackage{natbib}
\usepackage{url}
\usepackage{dcolumn}
\usepackage{multirow}
\usepackage{color}
\usepackage{cancel, comment}
\usepackage{soul}
\usepackage[normalem]{ulem}
\usepackage{amsfonts,amssymb,amsmath, txfonts}
\usepackage{graphicx,epsfig}
\usepackage{psfrag}
\usepackage{hyperref}
\usepackage{mathtools}
\usepackage{enumitem}
\usepackage{float}
\usepackage[dvipsnames]{xcolor}
\usepackage{xcolor}
\hypersetup{linktoc=all,
    colorlinks=true, linkcolor={brightpink},
    citecolor={blue}, urlcolor={blue}
}
\definecolor{rosy}{RGB}{230,235,252}
\definecolor{myframetitle}{RGB}{90,89,170}
\definecolor{myblocktitle}{RGB}{140,185,249}
\definecolor{mytitle}{RGB}{10,80,26}

\definecolor{darkgreen}{RGB}{27,130,45}
\definecolor{darkblue}{rgb}{0,0,0.3}
\definecolor{darkred}{rgb}{0.7,0,0}

\definecolor{light gray}{RGB}{220,220,220}
\definecolor{dark purple}{RGB}{108,0,217}
\definecolor{pink}{RGB}{190,20,100}
\definecolor{orang}{RGB}{193,63,0}
\definecolor{green}{RGB}{11,98,17}
\definecolor{darkpink}{RGB}{153,0,76}
\definecolor{bluegreen}{RGB}{0,102,102}
\definecolor{greenlagan}{RGB}{0,102,0}
\definecolor{redgreen}{RGB}{102,102,0}
\definecolor{Redgreen}{RGB}{153,76,0}
\definecolor{vividviolet}{rgb}{0.62, 0.0, 1.0}
\definecolor{amaranth}{rgb}{0.9, 0.17, 0.31}
\definecolor{palatinateblue}{rgb}{0.15, 0.23, 0.89}
\definecolor{brightpink}{rgb}{1.0, 0.0, 0.5}
\definecolor{cornflowerblue}{rgb}{0.39, 0.58, 0.93}
\definecolor{deepcarminepink}{rgb}{0.94, 0.19, 0.22}
\definecolor{radicalred}{rgb}{1.0, 0.21, 0.37}

\def\H0{{\text{H}\hspace*{-2.05mm}\text{H} 0\hspace*{-1.35mm}0\ }}

\def\be{\begin{equation}}
\def\ee{\end{equation}}
\def\beq{\begin{equation}}
\def\eeq{\end{equation}}
\def\bea{\begin{eqnarray}}
\def\eea{\end{eqnarray}}
\newcommand{\dd}{\textrm{d}}

\begin{document}

\title{Does DESI 2024 Confirm $\Lambda$CDM?}
\author{E. \'O Colg\'ain}
\affiliation{Atlantic Technological University, Ash Lane, Sligo, Ireland}
\author{M. G. Dainotti }
\affiliation{Division of Science, National Astronomical Observatory of Japan, 2-21-1 Osawa, Mitaka, 181-8588 Tokyo, Japan}  
\affiliation{The Graduate University for Advanced Studies (SOKENDAI), Shonankokusaimura, Hayama, Miura District, Kanagawa 240-0115,
Japan}
\affiliation{Space Science Institute, Boulder, CO, USA
}
\author{S. Capozziello}
\affiliation{Dipartimento di Fisica "E. Pancini" , Universit\'a degli Studi di  Napoli "Federico II"\\
                Complesso Univ. Monte S. Angelo, Via Cinthia 9
                80126, Napoli, Italy}
\affiliation{Scuola Superiore Meridionale, 
                Largo S. Marcellino 10,
                80138, Napoli, Italy}
\affiliation{Istituto Nazionale di Fisica Nucleare (INFN), Sez. di Napoli,
                Complesso Univ. Monte S. Angelo, Via Cinthia 9,
                80126, Napoli, Italy}

\author{S. Pourojaghi}
\affiliation{Department of Physics, Bu-Ali Sina University, Hamedan 65178, 016016, Iran}
\affiliation{School of Physics, Institute for Research in Fundamental Sciences (IPM),\\ P.O.Box 19395-5531, Tehran, Iran}
\author{M. M. Sheikh-Jabbari}
\affiliation{School of Physics, Institute for Research in Fundamental Sciences (IPM),\\ P.O.Box 19395-5531, Tehran, Iran}

\author{D. Stojkovic}
\affiliation{HEPCOS, Department  of  Physics,  SUNY  at  Buffalo,  Buffalo,  NY  14260-1500, USA}

\begin{abstract}
We demonstrate that a $\sim 2 \sigma$ discrepancy with the Planck-$\Lambda$CDM cosmology in DESI Luminous Red Galaxy (LRG) data in the redshift range $0.4 < z < 0.6$ with effective redshift $z_{\textrm{eff}} = 0.51$ translates into an unexpectedly large $\Omega_m$ value, $\Omega_m = 0.67^{+0.18}_{-0.17}$. We independently confirm that this anomaly drives the preference for $w_0 > -1$ in DESI data \textit{alone} confronted to the $w_0 w_a$CDM model. 
Given that LRG data at $z_{\textrm{eff}} = 0.51$ is at odds with Type Ia supernovae in overlapping redshifts, we expect that this anomaly will decrease in statistical significance with future DESI data releases leaving an increasing $\Omega_m$ trend with effective redshift at higher redshifts. We estimate the current significance of the latter in DESI data at $\sim 1.8 \sigma$ and comment on how it dovetails with independent observations. It is imperative to understand what makes DESI LRG data at $z_{\textrm{eff}} = 0.51$ an outlier when it comes to $\Omega_m$ determinations.

\end{abstract}

\maketitle

\section{Introduction}
Cosmological tensions \cite{DiValentino:2021izs,Perivolaropoulos:2021jda, Abdalla:2022yfr} and pre-calculus demand that we see changes in the $\Lambda$CDM model parameters when confronted to data in different redshift bins \cite{Krishnan:2020vaf, Krishnan:2022fzz} provided observational systematics are not at work. Conversely, if one recovers the same $\Lambda$CDM parameters within the errors when one confronts the model to data at different redshifts, then tensions point to systematics in data sets rather than a physical effect. In practice, testing all redshift ranges is not feasible, but we should endeavour to test as many redshift ranges as possible. The community is in the process of testing the consistency of the (flat) $\Lambda$CDM model. See \cite{Akarsu:2024qiq} for an overview of results. 

Recently, the Dark Energy Spectroscopic Instrument (DESI) collaboration has released its first round of cosmological constraints based on baryon acoustic oscillations (BAO) \cite{SDSS:2005xqv, 2dFGRS:2005yhx}, providing hints of evolution in the dark energy (DE) sector \cite{DESI:2024uvr, DESI:2024lzq, DESI:2024mwx}. Consider the $w_0w_a$CDM  model with redshift dependent DE equation of state, $w(z) = w_0 + (1-a) w_a = w_0 + z/(1+z) w_a$, where $a$ is the scale factor, $z$ is the redshift, and $(w_0, w_a) = (-1, 0)$ \cite{Chevallier:2000qy, Linder:2002et} recovers $\Lambda$CDM. In this context, DESI reports a preference for $w_0 > -1$ in DESI data alone, which is driven by the Luminous Red Galaxy (LRG) sample with effective redshift $z_{\textrm{eff}} = 0.51$. When DESI BAO data is combined with Planck CMB data \cite{Planck:2018vyg} and Type Ia supernovae (SNe), including Pantheon+ \cite{Brout:2022vxf}, Union3 \cite{Rubin:2023ovl} and DES \cite{DES:2024jxu}, the dynamical DE signal is driven by SNe preferring larger $\Omega_m$ values and primarily $z_{\textrm{eff}} = 0.71$ LRG preferring smaller $\Omega_m$ values, respectively, relative to Planck. 

While nothing precludes folding different data sets together and confronting them to the the $w_0w_a$CDM  model, care is required with physics. In cosmology, physics demands that observables are singing from the same hymn sheet. In short, late-time accelerated expansion is only believable if CMB, BAO, SNe, and many other data sets agree on its existence. Applying this argument to the current context, DESI claims of dynamical DE are only believable if BAO and SNe show consistent deviations from the $\Lambda$CDM model in similar redshift ranges. If they contradict each other, then statistical fluctuations, systematics, or evolutionary effects with redshifts not properly treated, are at play somewhere. 

In this letter we take a closer look at the implications of DESI data for the $\Lambda$CDM model on the grounds that if $w_0w_a$CDM  model is preferred over $\Lambda$CDM model, then this must be evident in changes or evolution of the $\Lambda$CDM paramaters with binned or effective redshift. Through this exercise, we find that $z_{\textrm{eff}} = 0.51$ LRG data, the most obvious outlier, prefers larger $\Omega_m \sim 0.65$ and $z_{\textrm{eff}} = 0.71$ LRG prefers smaller $\Omega_m \sim 0.2$ values of matter density. In  \textit{DESI data alone}, we confirm that $z_{\textrm{eff}} = 0.51$ LRG data is responsible for the preference for $w_0 > -1$ in the the $w_0w_a$CDM  model. Given that no other cosmological probe has a preference for $\Omega_m \sim 0.65$ at $z_{\textrm{eff}} = 0.51$, in particular SNe do not, we argue that a statistical fluctuation and/or systematics are at play either in DESI LRG or SNe data. Finally, we speculate that an increasing $\Omega_m$ trend with effective redshift in DESI data, driven in part by $z_{\textrm{eff}} = 0.71$ LRG, is physical by
invoking similar observations in independent probes.   

\section{Flat $\Lambda$CDM}
In this section we look for signatures of varying $\Lambda$CDM cosmological parameters in DESI data \cite{DESI:2024mwx}. The motivation is cosmological tensions \cite{DiValentino:2021izs,Perivolaropoulos:2021jda, Abdalla:2022yfr}, where it is necessary to diagnose potential shortcomings of the model by performing consistency checks \cite{Akarsu:2024qiq}. 

\subsection{Markov Chain Monte Carlo}
In Table~\ref{tab:DESI_OM_MCMC} we employ MCMC \cite{Foreman-Mackey:2012any} to determine 68\% credible intervals for $\Lambda$CDM parameters from anisotropic BAO constraints in Table~1 of \cite{DESI:2024mwx}. The observables include LRG, emission line galaxies (ELG) and the Lyman-$\alpha$ forest (Ly$\alpha$ QSO). The justification for focusing largely on anisotropic BAO is that, as is evident from Fig.~2 \cite{DESI:2024mwx}, isotropic BAO from the bright galaxies (BGS) and QSO cannot determine $\Omega_m$, and isotropic BAO only constrains posteriors to curves in the $(H_0, \Omega_m)$-plane.
Throughout we incorporate the correlation $r$ by constructing a $2 \times 2$ covariance matrix between $D_M/r_d$ and $D_H/r_d$ constraints at each effective redshift, where $r_d$ denotes the radius of the sound horizon at the baryon drag epoch and
\begin{equation}
    D_{M}(z) := \frac{c}{H_0} \int_0^z \ \frac{\dd z^{\prime}}{E(z^{\prime})} , \qquad D_{H}(z) := \frac{c}{H(z)}, 
\end{equation}
where $H_0$ is the Hubble constant, $H(z)= H_0 E(z)$ is the Hubble parameter and $c$ is  speed of light. Without external data or assumptions, it is impossible to separate $H_0$ from $r_d$, so we quote confidence intervals for the combination $H_0 r_d$ alongside $\Omega_m$. We adopt the usual uniform prior $\Omega_m \in [0,1]$ and a generous uniform prior for $H_0 r_d \in [0, 20000]$, neither of which impacts our results. We confirm this later by recovering the same results without MCMC and without imposing priors by directly converting DESI constraints on the ratio $(D_M/r_d)/(D_H/r_d)$ into the $\Lambda$CDM parameter $\Omega_m$.

\begin{figure}[htb]
\includegraphics[width=80mm]{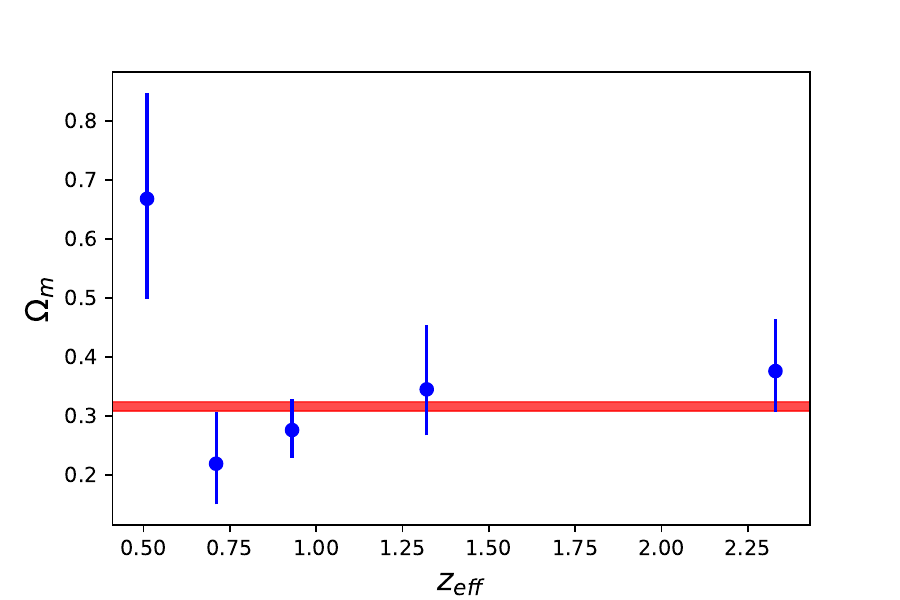}
\caption{$68 \%$ credible intervals for $\Omega_m$ versus effective redshift $z_{\textrm{eff}}$. Neglecting the DESI LRG constraint at $z_{\textrm{eff}} = 0.51$, the data shows good agreement with the corresponding Planck-$\Lambda$CDM confidence interval in red. The lower $\Omega_m$ in the full DESI dataset is primarily driven by $z_{\textrm{eff}} = 0.71$ LRG.}
\label{fig:OM_ZEFF} 
\end{figure}

From Table~\ref{tab:DESI_OM_MCMC}, there are essentially two features to highlight. First, the LRG data at $z_{\textrm{eff}} = 0.51$ leads to a surprisingly large value of $\Omega_m$ that is discrepant with the Planck value $\Omega_m = 0.315 \pm 0.007$ \cite{Planck:2018vyg} at the  $\sim 2.1 \sigma$ level. It is difficult to interpret this result, as it is also at odds with the Pantheon+ SNe constraint $\Omega_m = 0.334 \pm 0.018$ \cite{Brout:2022vxf} at the $\sim 2 \sigma$ level.\footnote{The relatively large LRG $\Omega_m$ errors mean that doubling the $\Omega_m$ error from CMB to SNe, while shifting the central value upwards, has little bearing on the statistical significance of the disagreement.} Note, SNe samples typically have low effective redshifts of $z_{\textrm{eff}} \sim 0.3$. Thus, in contrast to Planck versus DESI LRG, where one has observables at different redshifts, Pantheon+ versus DESI LRG, is a disagreement in an overlapping redshift range. While an unexpectedly large $\Omega_m$ may arise as a fluctuation in data, DESI has confirmed that the constraint in that redshift range is consistent with SDSS \cite{DESI:2024mwx}.\footnote{Between version 1 and version 2 (posted to the arXiv after this letter) of the arXiv submission \cite{DESI:2024mwx}, the DESI collaboration revised the paper to state that the $D_{M}/D_{H} (z_{\textrm{eff}}=0.51)$ constraint may be ``a mild ($\sim 2 \sigma$)
statistical fluctuation". The ratio $D_{M}/D_{H}$ directly constrains $\Omega_m$ in the $\Lambda$CDM model.} In particular, DESI reports a greater difference for the $z_{\textrm{eff}}=0.71$ LRG sample, but this can be attributed to the smaller errors, which are evident in Fig. \ref{fig:OM_ZEFF}. However, comparison to Fig. 5 of \cite{Colgain:2022nlb}, reveals that the inferred $\Omega_m$ value at $z_{\textrm{eff}} \sim 0.5$ has shifted upwards from $\Omega_m \sim 0.35$ in SDSS \cite{BOSS:2016wmc, eBOSS:2020yzd} to $\Omega_m \sim 0.65$ in DESI results, whereas the inferred $\Omega_m$ at $z_{\textrm{eff}} \sim 0.7$ has shifted downwards from $\Omega_m \sim 0.5$ in SDSS to $\Omega_m \sim 0.2$ in DESI results. Given these observations, it is difficult to preclude a statistical fluctuation either in SDSS or DESI data. Interestingly, DES also reports a $D_{M}/r_d$ constraint at $z_{\textrm{eff}} = 0.85$, which in agreement with DESI LRG at $z_{\textrm{eff}} = 0.71$ is $\sim 2 \sigma$ lower than the standard model \cite{DES:2024pwq}.

\begin{table}[htb]
    \centering
    \begin{tabular}{c|c|c|c}
    \rule{0pt}{3ex} tracer & $z_{\textrm{eff}}$ & $H_0 r_d$ [100 km/s] & $\Omega_m$ \\
    \hline
\rule{0pt}{3ex} LRG & $0.51$ & $88.3^{+5.7}_{-4.8}$ & $0.67^{+0.18}_{-0.17}$ \\
 \rule{0pt}{3ex} LRG & $0.71$ & $109.4^{+6.1}_{-6.1}$&  $0.219^{+0.087}_{-0.069}$ \\
\rule{0pt}{3ex} LRG+ELG & $0.93$ & $102.2^{+4.1}_{-4.0}$ & $0.276^{+0.053}_{-0.047}$ \\
\rule{0pt}{3ex} ELG & $1.32$ & $97.8^{+7.7}_{-8.0}$ & $0.345^{+0.11}_{-0.078}$ \\
\rule{0pt}{3ex} Ly$\alpha$ QSO & $2.33$ & $92.8^{+7.6}_{-7.5}$ & $0.375^{+0.088}_{-0.069}$ \\
    \end{tabular}
    \caption{ $68 \%$ credible intervals from DESI anisotropic BAO constraints at effective redshift $z_{\textrm{eff}}$.}
    \label{tab:DESI_OM_MCMC}
\end{table}

The second feature of interest is an increasing $\Omega_m$/decreasing $H_0 r_d$ with effective redshift beyond the anomalous LRG constraint at $z_{\textrm{eff}}=0.51$. The trend is evident in Fig.~\ref{fig:OM_ZEFF} and Table~\ref{tab:DESI_OM_MCMC}, where it is clearly driven by $z_{\textrm{eff}}=0.71$ LRG, since this provides the lowest $\Omega_m$ constraint. This trend is expected if observations of similar trends of decreasing $H_0$ and/or increasing $\Omega_m$ with redshift in strong lensing time delay, both lensed QSOs \cite{Wong:2019kwg, DES:2019fny, Millon:2019slk} and SNe \cite{Kelly:2023mgv, Pascale:2024qjr},\footnote{The claim is particularly clear from Fig.~5 of  \cite{Pascale:2024qjr} (see also \cite{Li:2024elb}), where it is evident that the error bars of $H_0$ determined from SN Refsdal and SN H0pe do not overlap, placing the disagreement at $\sim 1.5 \sigma$. Note that SN H0pe has a lens redshift of $z = 0.35$, whereas the lens redshift of SN Refsdal is $z = 0.54$, thereby making Fig.~5 of \cite{Pascale:2024qjr} consistent with the descending trend of $H_0$ with lens redshift reported originally in Wong et al. \cite{Wong:2019kwg} (see appendix).} Type Ia SNe \cite{Dainotti:2021pqg, Colgain:2022nlb, Colgain:2022rxy, Jia:2022ycc, Pasten:2023rpc, Malekjani:2023ple} (see also \cite{Hu:2022kes,Wagner:2022etu, Dainotti:2023yrk} for related discussions), observational Hubble data (OHD) \cite{Dainotti:2022bzg, Colgain:2022rxy}, combinations of OHD and SNe \cite{Krishnan:2020obg, Colgain:2022rxy}, GRBs \cite{Dainotti:2022rea}, and standardsiable QSOs \cite{Risaliti:2018reu, Lusso:2020pdb, Colgain:2022nlb, Colgain:2022rxy,Dainotti:2022rfz,Dainotti:2023cpn,Pourojaghi:2022zrh,Bargiacchi:2023jse} hold up\footnote{The original papers \cite{Risaliti:2018reu, Lusso:2020pdb} highlight a high redshift deviation from Planck-$\Lambda$CDM behaviour, whereby the luminosity distance $D_{L}(z)$ preferred by the QSOs is smaller than the Planck-$\Lambda$CDM model. This translates into a discrepancy in $\Omega_m$ in the $\Lambda$CDM model.}. These observations support a late Universe resolution to $\Lambda$CDM tensions \cite{Vagnozzi:2023nrq}. 

As an aside, there is a growing body of work in the literature either questioning \cite{Khadka:2020vlh, Khadka:2020tlm, Singal:2022nto, Petrosian:2022tlp, Zajacek:2023qjm} or improving the Risaliti-Lusso standardisable QSO prescription \cite{Dainotti:2022rfz, Dainotti:2024aha, Dainotti:2024bth}. Nevertheless, despite the corrections imposed on the QSO data, residual evolution of the $\Omega_m$ parameter is still reported \cite{Dainotti:2024aha, Dainotti:2024bth, Lenart:2022nip}. Moreover, DES SNe have a higher effective redshift \cite{DES:2024jxu}, so the larger $\Omega_m$ value is consistent with
these observations (see also \cite{Colgain:2024ksa}). Going beyond established probes such as SNe,  QSOs \cite{Risaliti:2018reu, Lusso:2020pdb} and GRBs \cite{Dainotti:2008vw, Demianski:2016zxi, Demianski:2016dsa, Srinivasaragavan:2020isz,Dainotti:2022ked, Khadka:2021vqa, Alfano:2024ukk} calibrated by SNe, may also return larger values of $\Omega_m$ at higher redshifts. \footnote{This could be mundanely due to large scatter in the data and not due to any variation in the $\Lambda$CDM parameters. One can separate these two possibilities by comparing to SNe in overlapping redshift regimes \cite{Colgain:2022nlb}.} Returning to the main point, if not for the LRG $\Omega_m$ constraint at $z_{\textrm{eff}} = 0.51$, one could add the four highest redshift DESI constraints from Fig.~\ref{fig:OM_ZEFF} to the corroborating support for the increasing $\Omega_m$/decreasing $H_0$ trend.  

\begin{table}[htb]
    \centering
    \begin{tabular}{c|c|c}
    \rule{0pt}{3ex} $z$ & $H_0 r_d$ [100 km/s] & $\Omega_m$ \\
    \hline
     \rule{0pt}{3ex} $0.1 < z < 0.6$ & $96.1^{+3.8}_{-3.9}$&  $0.459^{+0.12}_{-0.098}$ \\
 \rule{0pt}{3ex} $0.6 < z < 1.1$ & $106.8^{+3.1}_{-3.1}$ &  $0.231^{+0.036}_{-0.033}$ \\
    \rule{0pt}{3ex}  $1.1 < z < 4.16$ & $98.6^{+4.3}_{-4.3}$ & $0.324^{+0.044}_{-0.038}$ \\
    \end{tabular}
    \caption{$68 \%$ credible intervals from DESI anisotropic and isotropic BAO binned by redshift. From bin 1 to bin 2 $\Omega_m$ shifts downwards by $\sim 2.2 \sigma$ and from bin 2 to bin 3 $\Omega_m$ shifts upwards by $ \sim 1.8 \sigma$.}
    \label{tab:DESI_OM_MCMC_low_high}
\end{table}

To assess the statistical significance of the increasing $\Omega_m$/decreasing $H_0 r_d$ trend at higher redshifts, as well as the impact of the DESI constraint at $z_{\textrm{eff}} = 0.51$, we introduce the remaining two isotropic BAO constraints from Table 1 of the DESI paper \cite{DESI:2024mwx}. The corresponding $68 \%$ credible intervals are shown in Table~\ref{tab:DESI_OM_MCMC_low_high} in redshift bins, where it is evident that $\Omega_m$ starts off high at lower redshifts, decreases at intermediate redshifts, before increasing again at larger redshifts. It has been shown in \cite{Colgain:2024mtg} that this is also the outcome of mapping the $w_0 w_a$CDM cosmologies preferred by DESI+CMB+SNe back into the flat $\Lambda$CDM parameter $\Omega_m$, but here our observations only concern DESI BAO. Barring the first bin $0.1 < z < 0.6$, which we include for completeness in Table~\ref{tab:DESI_OM_MCMC_low_high}, we are primarily interested in the increase in $\Omega_m$ with redshift in Fig. \ref{fig:OM_ZEFF}. The point of Table~\ref{tab:DESI_OM_MCMC_low_high} is to split the 4 remaining constraints into a low and high redshift subsample to give a first approximation estimate of the statistical significance of the increase in $\Omega_m$ values at higher redshifts. This motivates the choice of bins. The statistical significance of the shifts in the $\Omega_m$ parameter are $\sim 2.2 \sigma$ lower and $\sim 1.8 \sigma$ higher, respectively. {We explored replacing DESI constraints with SDSS constraints \cite{eBOSS:2020yzd} while avoiding data overlap, but failed to find shifts as pronounced.} As DESI is designed to produce a much larger number of constraints at a wide range of redshifts \cite{DESI:2016fyo}, it will be important to revisit these two features with future data releases. Note, at $\sim 2 \sigma$ one can of course ignore these results, but it is well established that matter density is approximately $30\%$ in the $\Lambda$CDM Universe. This now comes with the caveat that there is a corner of the Universe probed by DESI where matter density is $65\%$; the inconsistency is striking.     

Before moving on, we revisit Fig.~3 of \cite{DESI:2024mwx} to comment on the discrepancies with Planck, $\Omega_m = 0.315 \pm 0.007$ \cite{Planck:2018vyg}, Pantheon+,  $\Omega_m = 0.334 \pm 0.018$ \cite{Brout:2022vxf}, Union3, $\Omega_m = 0.356^{+0.028}_{-0.026}$ \cite{Rubin:2023ovl}, and DES, $\Omega_m = 0.352 \pm 0.017$ \cite{DES:2024jxu}, data sets. Once again, we use all the data, both anisotropic and isotropic BAO constraints. For all constraints in Table 1 of \cite{DESI:2024mwx} we find $\Omega_m = 0.290^{+0.015}_{-0.014}$. This agrees well with our best fit or maximum likelihood estimator (MLE) of $\Omega_m = 0.289$. Note, the central value we find is slightly lower than Table 3 of the DESI paper \cite{DESI:2024mwx} \footnote{We have confirmed that the $\sim 0.3 \sigma$ shift is due to a rounding up of the $z_{\textrm{eff}}$ redshifts from version 2 (3 decimal places) to version 1 (2 decimal places) of the DESI arXiv submission \cite{DESI:2024mwx}.}. This sets the disagreement with Planck, Pantheon+, Union3 and DES at $\sim 1.5 \sigma$, $\sim 1.9 \sigma$, $\sim 2.2 \sigma$ and $\sim 2.7 \sigma$ , respectively. While all disagreements are below a $3 \sigma$ threshold, they are serious enough that further study is warranted. It is fitting to recall the claim originally made in \cite{Colgain:2022nlb} that the fitting parameter $\Omega_m$ is not a constant and evolves, or more precisely increases, with effective redshift. 

Finally, we remove the LRG contraint at $z_{\textrm{eff}} = 0.51$. Doing so, we expect the $\Omega_m$ value to decrease,  and it does to $\Omega_m = 0.287^{+0.016}_{-0.015}$ with an MLE of $\Omega_m = 0.286$. Overall, the shift downwards is not pronounced, and the error inflates, but this increases the tension with DES to $\sim 2.8 \sigma$. Evidently, the working assumption that $\Omega_m$ is a constant in the $\Lambda$CDM Universe is under pressure at close to $3 \sigma$. This needs to be put in context. Given $H_0$ tension, a problem in background cosmology,\footnote{$H_0 = H(z=0)$ is determined by extrapolating the Hubble parameter, itself a solution to the Einstein equations, from low redshifts to $z = 0$.} and $S_8 = \sigma_8 \sqrt{\Omega_m/0.3}$ tension, a seemingly perturbative problem, it is important to recognise that both these problems can be related provided we find a discrepancy in $\Omega_m$ or ``$\Omega_m$ tension" \cite{Akarsu:2024qiq}. The reason being that $H_0$ is correlated with $\Omega_m$ at the background level in the late Universe and $S_8$ clearly depends on $\Omega_m$. If $\Omega_m$ is not a constant in the $\Lambda$CDM model, then $H_0$ and $S_8$ tensions are symptoms of the same underlying problem.  

\subsection{A second look}
In this section we adopt complementary methodology following \cite{Colgain:2022nlb}. This provides a check independent of MCMC on the results from anisotropic BAO in Table~\ref{tab:DESI_OM_MCMC}. The basic idea is to employ the  fact that the ratio 
\begin{equation}
\label{eq:DMDH}
    \frac{D_M/r_d}{D_H/r_d} = E(z) \int_0^z \frac{1}{E(z^{\prime})} \dd z^{\prime}, 
\end{equation}
only depends on $\Omega_m$ in the $\Lambda$CDM model for which $E(z) = \sqrt{1-\Omega_m + \Omega_m (1+z)^3}$. We remark that this ratio is the same as the combination of Alcock-Paczynski-like dilation parameters $\alpha_{\textrm{AP}} \equiv \alpha_{\parallel}/\alpha_{\perp}$ \cite{Alcock:1979mp}.

\begin{figure}
    \includegraphics[width=80mm]{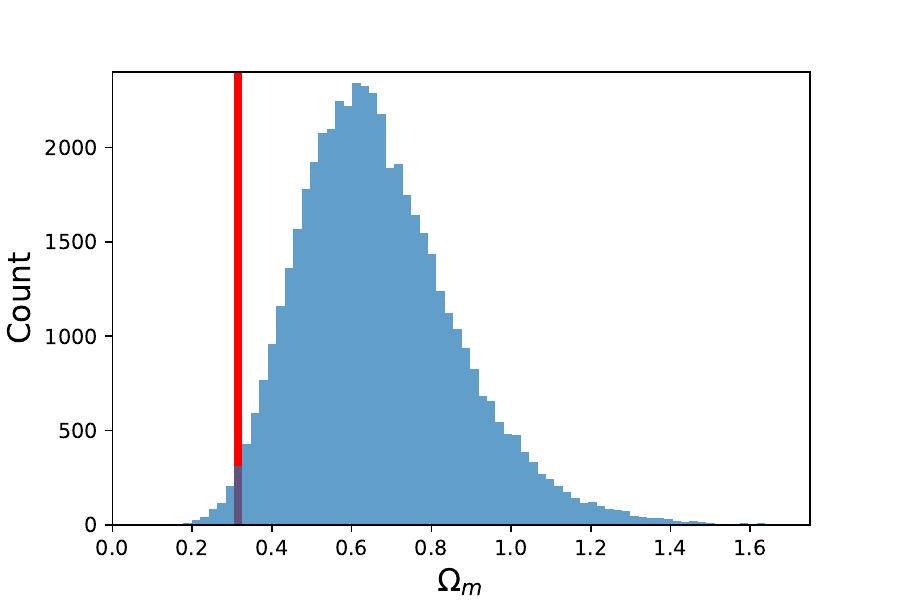}
    \caption{Distribution of 50,000 $\Omega_m$ values at $z_{\textrm{eff}}=0.51$ alongside Planck constraints $\Omega_m = 0.315 \pm 0.007$ in red. The probability of finding an $\Omega_m$ value within the Planck $1 \sigma$ confidence interval or lower is $p = 0.015$ corresponding to $2.2 \sigma.$ }
    \label{fig:LRG_OM}
\end{figure}

Eq.~(\ref{eq:DMDH}) allows one to numerically solve for $\Omega_m$.\footnote{We check solutions to a precision of $10^{-10}$ by reinserting the solution in \eqref{eq:DMDH}.} In contrast to the previous section, here we impose no bound (prior) on $\Omega_m$. To incorporate the errors, one employs the $68 \%$ confidence intervals and the correlation $r$ to construct a $2 \times 2$ covariance matrix for each effective redshift. From the covariance matrix, one randomly generates a large number (approx. 10,000) of $(D_{M}/r_d, D_{H}/r_d)$ pairs in a multivariate normal distribution. For each pair $(D_{M}/r_d, D_{H}/r_d)$, one solves the right hand side of (\ref{eq:DMDH}) for $\Omega_m$. This builds up a distribution of $\Omega_m$ values from which one extracts the mean and the $68 \%$ confidence intervals in Table~\ref{tab:DESI_OM_new_method}. More precisely, we determine the $68 \%$ confidence interval by identifying the $15.9$ and $84.1$ percentiles. Throughout we assume that the $\Omega_m$ distribution is Gaussian, but as is evident from the errors in Table~\ref{tab:DESI_OM_new_method}, and Fig.~\ref{fig:LRG_OM}, this is not a bad approximation \footnote{In principle, for non-Gaussian posteriors one may fit a different distribution and this has a bearing on the errors \cite{Dainotti:2023bwq}. Here we expect any difference to be small.}. The main take-away is that Table~\ref{tab:DESI_OM_MCMC} and Table \ref{tab:DESI_OM_new_method} show excellent agreement in central values and errors. From the outset, this may not have been expected as \eqref{eq:DMDH} removes all dependence on $H_0 r_d$, so one is only constraining $\Omega_m$. Smaller errors may have been expected, but this was not the case. 

\begin{table}[htb]
    \centering
    \begin{tabular}{c|c|c}
    \rule{0pt}{3ex} tracer & $z_{\textrm{eff}}$ & $\Omega_m$ \\
    \hline
\rule{0pt}{3ex} LRG & $0.51$ & $0.65^{+0.21}_{-0.17}$ \\
\rule{0pt}{3ex} LRG & $0.71$ & $0.210^{+0.082}_{-0.068}$ \\
    \rule{0pt}{3ex} LRG+ELG & $0.93$ & $0.268^{+0.052}_{-0.045}$ \\
 \rule{0pt}{3ex} ELG & $1.32$ & $0.330^{+0.10}_{-0.076}$ \\
    \rule{0pt}{3ex} Lyman-$\alpha$ QSO & $2.33$ & $0.362^{+0.084}_{-0.066}$ \\
    \end{tabular}
    \caption{ $68 \%$ $\Omega_m$ confidence intervals from DESI anisotropic BAO constraints at effective redshift $z_{\textrm{eff}}$ using Eq. (\ref{eq:DMDH}).}
    \label{tab:DESI_OM_new_method}
\end{table}

Using Eq. (\ref{eq:DMDH}) we revisit the anomaly in LRG data evident in Fig.~\ref{fig:LRG_OM}. We increase the number of configurations to 50,000, where we find a probability of $p = 0.0147$ of getting a value lower than $\Omega_m = 0.322$, the upper bound on the Planck $68 \%$ confidence intervals. This places the disagreement in $\Omega_m$ at no less than $2.2 \sigma$.

\section{$w_0w_a$CDM  model}
From Fig.~\ref{fig:OM_ZEFF} and Fig.~\ref{fig:LRG_OM} it is clear that LRG data at $z_{\textrm{eff}} = 0.51$ disagree with the Planck-$\Lambda$CDM model at the $\gtrsim 2 \sigma$ level. Moreover, as explained earlier, this constraint on $\Omega_m$ also disagrees with Pantheon+ \cite{Brout:2022vxf} at the $\sim 2 \sigma$ level, despite Type Ia SNe being most sensitive to comparable redshift ranges. As emphasised, this disagreement between DESI and Pantheon+ SNe, and more generally any SNe data set, since all closely agree on $\Omega_m \sim 0.3$ at lower redshifts, warrants further exploration. That point aside, while DESI constraints on the $w$CDM model from Table~3 of \cite{DESI:2024mwx} are consistent with a cosmological constant interpretation ($w=-1$) within $1 \sigma$, a fit of the $w_0w_a$CDM  model leads to a different conclusion that $w = w_0 > -1$ is preferred at in excess of $2 \sigma$. This overestimates the significance as the $w_a$ parameter is poorly constrained and this necessitates combining DESI with CMB and SNe.

\begin{figure}[htb]
    \includegraphics[width=80mm]{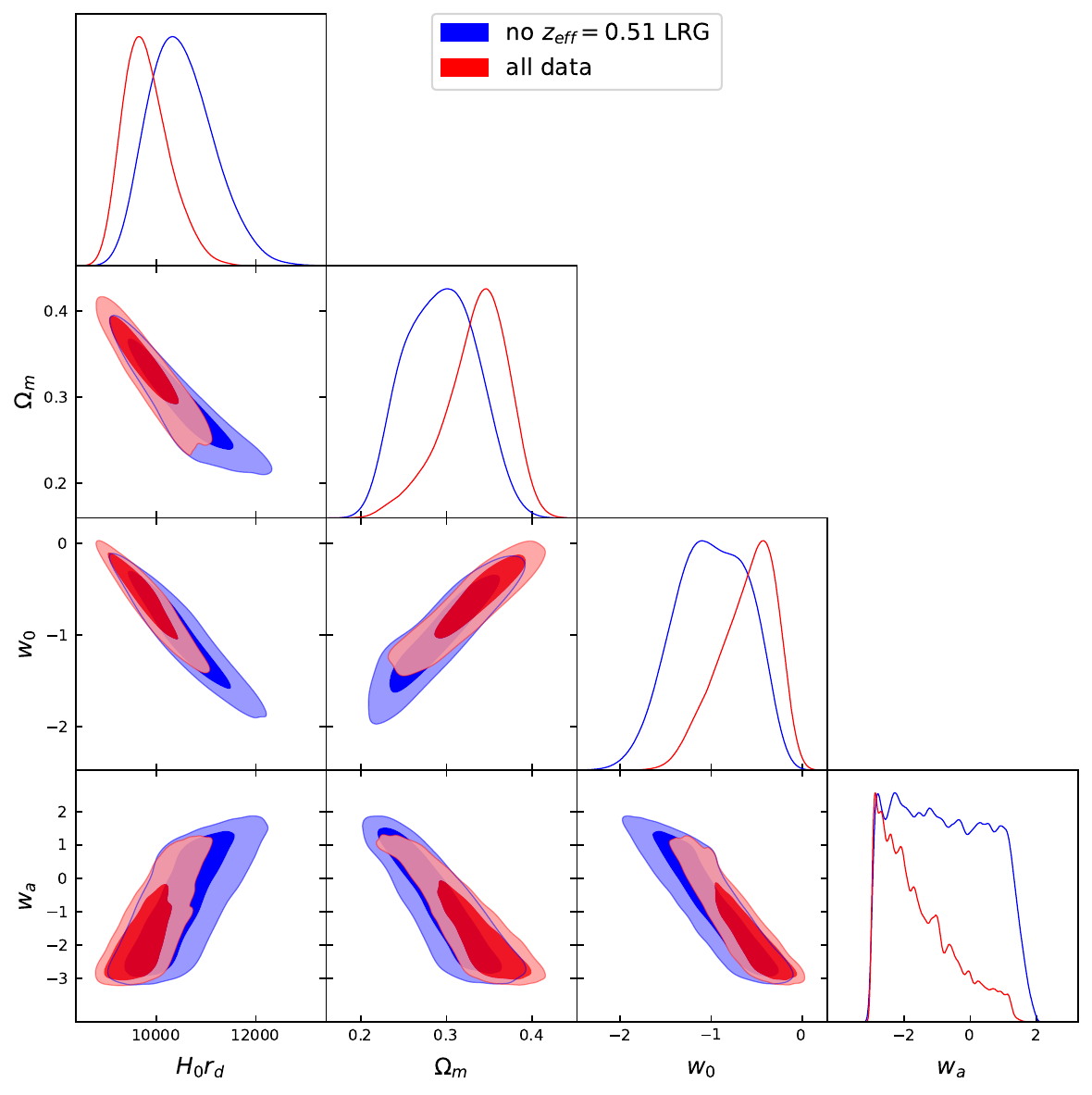}
    \caption{$w_0w_a$CDM  model with and without LRG data at $z_{\textrm{eff}} = 0.51$. Any hint of dynamical DE, in particular $w_0 > -1$, in DESI data is driven by LRG data. We acknowledge use of \textit{GetDist} \cite{Lewis:2019xzd}.}
    \label{fig:CPL}
\end{figure}

We demonstrate here that this deviation from $\Lambda$CDM is driven by LRG data at $z_{\textrm{eff}} = 0.51$. This is acknowledged in the DESI paper \cite{DESI:2024mwx}, so our goal here is to provide independent confirmation. We employ MCMC with uniform DESI priors, $w_0 \in [-3, 1], w_a \in [-3, 2]$ and $w_0+w_a <0$. The key point is that if data prefers the $w_0w_a$CDM  model over $\Lambda$CDM, then changes in the flat $\Lambda$CDM parameter $\Omega_m$ with effective (or binned) redshift must be evident in the $\Lambda$CDM model. See \cite{Colgain:2024mtg} for an explicit demonstration of how one maps a $\Lambda$CDM deviation in the $w_0 w_a$CDM model back into a non-constant $\Omega_m$ in the minimal model. In Fig.~\ref{fig:CPL} we confirm that removing the most striking deviation from Planck-$\Lambda$CDM behaviour in Fig.~\ref{fig:OM_ZEFF} returns $w_0$ to $w_0 \sim -1$. We remark that one can clearly see the effect of DESI priors, especially the $w_a$ and $w_0 + w_a$ priors, on confidence intervals in the $(w_0, w_a)$-plane; $w_a$ is poorly constrained without external data.  

Concretely, we find $w_0 = -0.56^{+0.27}_{-0.38}$ with the anomalous LRG data and $w_0 =  -0.98^{+0.43}_{-0.42}$ without. Evidently, $w_a$ is pushed to large negative values that are beyond the priors to accommodate $w_0 > -1$ {(see also comments in \cite{Cortes:2024lgw}). Unsurprisingly, it has recently been shown that relaxing the $w_a$ priors drives $w_0$ to a larger value, $w_0 \sim 1$ \cite{Wang:2024rjd}, which is inconsistent with late-time accelerated expansion requiring $w(z) < -\frac{1}{3}$. This underscores that the $w_0 w_a$CDM model is poorly constrained with DESI data alone. Nevertheless}, neglecting the LRG data, DESI shows excellent agreement with $\Lambda$CDM. One concludes that any hint of dynamical DE in DESI data is driven by the LRG data. Note, all our analysis rests upon DESI BAO data alone. When DESI is combined with CMB and SNe, it has been reported in the literature \cite{Wang:2024pui, Chudaykin:2024gol} that  $z_{\textrm{eff}} = 0.71$ plays a more dominant role driving the signal of dynamical DE in BAO+CMB+SNe combinations. This is easily understood as follows. If BAO, CMB and SNe all agree on $\Omega_m$, then one does not expect to see a $\Lambda$CDM deviation in the $w_0 w_a$CDM model. Thus, this signal must be driven by the $\Omega_m$ discrepancies evident in Fig. 3 of \cite{DESI:2024mwx}. Noting that DESI BAO leads to the smallest $\Omega_m$ value, it is clear from Fig. \ref{fig:OM_ZEFF} that this low $\Omega_m$ value is driven primarily by $z_{\textrm{eff}} = 0.71$ LRG data.     

\section{Discussion}
With its own data DESI has reported hints \cite{DESI:2024mwx} of a preference for an evolving DE equation of state in the $w_0w_a$CDM  model. As explained in the original paper, and confirmed here in Fig.~\ref{fig:OM_ZEFF} and Fig.~\ref{fig:CPL}, the preference for $w_0 > -1$ can be traced to the LRG data with $z_{\textrm{eff}} = 0.51$. While the tension with Planck-$\Lambda$CDM is only at the $\sim 2 \sigma$ level, translated into more familiar $\Lambda$CDM parameters, DESI LRG data at $z_{\textrm{eff}} = 0.51$ favours a $\Omega_m \sim 0.65$ best fit. This places DESI LRG constraints at odds not only with Planck, but also with multiple Type Ia SNe samples \cite{Brout:2022vxf, Rubin:2023ovl, DES:2024jxu}, where it should be stressed that the effective redshift is not dissimilar; SNe, being numerous at lower redshift, are more sensitive to the lowest redshift bins probed by BGS and LRG tracers. Given that SDSS and DESI constraints agree \cite{DESI:2024mwx}, it should come as no surprise that SDSS also prefers larger $\Omega_m$ in the same range \cite{Colgain:2022nlb} (see Fig.~5), but admittedly not quite as large. Shifts in the $\Omega_m$ values between SDSS and DESI constraints are evident at lower redshifts, which systematics aside hint at statistical fluctuations in the data. A revision in the DESI paper \cite{DESI:2024mwx} now makes it clearer that the DESI result for $D_M/D_H$ at $z_{\textrm{eff}} = 0.51$, which equates to the anomalously large $\Omega_m$ reported here, appears consistent with a statistical fluctuation.     

Arguably the great success of the $\Lambda$CDM model is that CMB, BAO and SNe agree that the Universe is 70\% dark energy, $\Omega_m \sim 0.3$. In cosmology, physics must be supported by multiple observables, otherwise one is looking at statistical fluctuations or observational systematics, here either in LRG constraints at $z_{\textrm{eff}} = 0.51$ or in 3 SNe samples.\footnote{These samples are not fully independent, so it is not implausible that systematics persist in SNe. Pantheon+ and Union3 share $\sim 1360$ SNe, while Pantheon+ and DES share 196 SNe.} This is the main take-away message from this letter. Of course, one can draw comparisons in the behaviour of the $w_0w_a$CDM  model, but care is required to make sure one is seeing consistent deviations from $\Lambda$CDM in independent observables in the same redshift ranges, otherwise claims of new physics are not compelling, no matter what interest is generated \cite{Tada:2024znt, Gu:2024jhl, Wang:2024qan, Wang:2024hks, Luongo:2024fww, Yin:2024hba}. 

As is evident from Table~\ref{tab:DESI_OM_MCMC_low_high}, even in DESI data confronted to the $\Lambda$CDM model, shifts of $\sim 2 \sigma$ are evident in $\Omega_m$ as the data is binned. Concretely, $\Omega_m$ decreases with effective redshift before increasing again. $H_0 r_d$ is anti-correlated with $\Omega_m$, so it increases and decreases again. Historically, Type Ia SNe samples have shown similar trends \cite{Colgain:2019pck, Kazantzidis:2020xta}, but this is not overly surprising given that large SNe samples are typically compiled from different surveys. As the number of surveys grows, observational systematics become a greater concern. Nevertheless, in recent years, the Pantheon+ sample has been demonstrated to largely return consistent $\Omega_m$ values \cite{Brout:2022vxf}, provided one neglects high redshift SNe \cite{Malekjani:2023ple}. In contrast, in DES SNe \cite{DES:2024jxu}, which is largely a sample from a single survey, the reported $\sim 2 \sigma$ dynamical dark energy signal translates into an $\Omega_m$ value that increases with redshift \cite{Colgain:2024ksa}. In the case of DESI, one is looking at a single survey, but multiple tracers.    

The big picture here is that persistent $\Lambda$CDM tensions, most notably discrepancies in $H_0$ and $S_8 = \sigma_8 \sqrt{\Omega_m/0.3}$ \cite{DiValentino:2021izs,Perivolaropoulos:2021jda, Abdalla:2022yfr}, assuming they are physical in origin, demand the $\Lambda$CDM model breaks down through cosmological (fitting) parameters that change with effective redshift \cite{Krishnan:2020vaf, Krishnan:2022fzz}.\footnote{One could argue that $S_8$ is a scale problem, e. g. \cite{Amon:2022azi, Preston:2023uup}, but this cannot resolve the statistically more significant $H_0$ tension problem.  Moreover, shifts in $S_8$ due to changes in scale appear too small to resolve $S_8$ tension \cite{Terasawa:2024agq}.} In support, observations of decreasing $H_0$/increasing $\Omega_m$ values with increasing effective redshift have been reported in a host of observables \cite{Wong:2019kwg, DES:2019fny, Millon:2019slk, Kelly:2023mgv, Pascale:2024qjr, Dainotti:2021pqg, Colgain:2022nlb, Colgain:2022rxy, Jia:2022ycc, Pasten:2023rpc, Malekjani:2023ple, Dainotti:2022bzg, Colgain:2022rxy, Krishnan:2020obg, Dainotti:2022bzg, Dainotti:2022rea, Risaliti:2018reu, Lusso:2020pdb, Pourojaghi:2022zrh}. If substantiated, these findings point to a breakdown of the $\Lambda$CDM model at the  \textit{background level} in the late Universe. Coincidentally, observations also exist to localise $S_8$ tension, an apparent problem \textit{at the perturbative level}, to the late Universe, $z \lesssim 2$ \cite{Esposito:2022plo, Adil:2023jtu, ACT:2023kun, Tutusaus:2023aux}. See \cite{Akarsu:2024qiq} for a review of the science case supporting redshift evolution of $\Lambda$CDM parameters.

The situation is finely balanced. DESI has started data releases and constraints are expected to improve in coming years \cite{DESI:2016fyo}. If the anomaly in LRG data at $z_{\textrm{eff}} = 0.51$ persists, then BAO and CMB+SNe, and potentially even CMB and SNe \cite{DES:2024jxu}, are at odds on $\Omega_m$. On the flip side, if improved data quality decreases the tension with Planck, Pantheon+ etc, it is plausible that the decreasing $\Omega_m$ (increasing $H_0 r_d$) trend at lower redshifts disappears leaving an increasing $\Omega_m$ (decreasing $H_0 r_d$) trend that is seen elsewhere. 

\textit{Note added:} Since this paper appeared on the arXiv, the DESI collaboration has released full-shape modelling results of data release 1 (DR1) \cite{DESI:2024hhd, DESI:2024jis} and updated BAO constraints for DR2 \cite{DESI:2025zgx}. In DR1 full-shape modelling, one can interpolate a constant $\Omega_m$ through all the constraints when the data is binned, but $\Omega_m$ increases with redshift across the bins with a mild statistical significance of $0.8 \sigma$. This currently corroborates our secondary claim in this letter, but the errors are too large to draw a firm conclusion. The same analysis as this letter has been repeated for DESI DR2 BAO \cite{Colgain:2025nzf}, where it is noted that fluctuations are still evident in LRG $\Omega_m$ constraints, but the general trend is towards greater consistency with constant $\Omega_m$ $\Lambda$CDM behaviour. The latter is also supported by DR1 full-shape modelling at current precision \cite{DESI:2024hhd, DESI:2024jis} and we can expect DESI full-shape modelling and BAO to further converge in future releases.

\section*{Acknowledgments}
We thank Santiago Avila, Anton Chudaykin, Bin Hu, Sunny Vagnozzi and Deng Wang for interesting discussions. DS is partially supported by the US National Science Foundation, under Grants No.
PHY-2014021 and PHY-2310363. MMShJ would like to acknowledge SarAmadan grant INSF No. 4026712 and ICTP HECAP section where this project was carried out. M.G. acknowledge the support of Division of Science at NAOJ. SP would like to acknowledge the support of the Iran National Science Foundation (INSF) for the research funding provided under project No. 4024802. This article/publication is based upon work from COST Action CA21136 – “Addressing observational tensions in cosmology with systematics and fundamental physics (CosmoVerse)”, supported by COST (European Cooperation in Science and Technology).


\begin{thebibliography}{99}

\bibitem{DiValentino:2021izs}
E.~Di Valentino, O.~Mena, S.~Pan, L.~Visinelli, W.~Yang, A.~Melchiorri, D.~F.~Mota, A.~G.~Riess and J.~Silk,
``In the realm of the Hubble tension: a review of solutions,''
Class. Quant. Grav. \textbf{38} (2021) no.15, 153001
[arXiv:2103.01183 [astro-ph.CO]].

\bibitem{Perivolaropoulos:2021jda}
L.~Perivolaropoulos and F.~Skara,
``Challenges for \ensuremath{\Lambda}CDM: An update,''
New Astron. Rev. \textbf{95} (2022), 101659
[arXiv:2105.05208 [astro-ph.CO]].

\bibitem{Abdalla:2022yfr}
E.~Abdalla, G.~Franco Abell\'an, A.~Aboubrahim, A.~Agnello, O.~Akarsu, Y.~Akrami, G.~Alestas, D.~Aloni, L.~Amendola and L.~A.~Anchordoqui, \textit{et al.}
``Cosmology intertwined: A review of the particle physics, astrophysics, and cosmology associated with the cosmological tensions and anomalies,''
JHEAp \textbf{34} (2022), 49-211
[arXiv:2203.06142 [astro-ph.CO]].

\bibitem{Krishnan:2020vaf}
C.~Krishnan, E.~\'O~Colg\'ain, M.~M.~Sheikh-Jabbari and T.~Yang,
``Running Hubble Tension and a H0 Diagnostic,''
Phys. Rev. D \textbf{103} (2021) no.10, 103509
[arXiv:2011.02858 [astro-ph.CO]].

\bibitem{Krishnan:2022fzz}
C.~Krishnan and R.~Mondol,
``$H_0$ as a Universal FLRW Diagnostic,''
[arXiv:2201.13384 [astro-ph.CO]].

\bibitem{Akarsu:2024qiq}
\"O.~Akarsu, E.~\'O~Colg\'ain, A.~A.~Sen and M.~M.~Sheikh-Jabbari,
``$\Lambda$CDM Tensions: Localising Missing Physics through Consistency Checks,''
\textit{Universe (2024) \textbf{10} no.8,  305}, 
[arXiv:2402.04767 [astro-ph.CO]].

\bibitem{SDSS:2005xqv}
D.~J.~Eisenstein \textit{et al.} [SDSS],
``Detection of the Baryon Acoustic Peak in the Large-Scale Correlation Function of SDSS Luminous Red Galaxies,''
Astrophys. J. \textbf{633} (2005), 560-574
[arXiv:astro-ph/0501171 [astro-ph]].

\bibitem{2dFGRS:2005yhx}
S.~Cole \textit{et al.} [2dFGRS],
``The 2dF Galaxy Redshift Survey: Power-spectrum analysis of the final dataset and cosmological implications,''
Mon. Not. Roy. Astron. Soc. \textbf{362} (2005), 505-534
[arXiv:astro-ph/0501174 [astro-ph]].

\bibitem{DESI:2024uvr}
A.~G.~Adame \textit{et al.} [DESI],
``DESI 2024 III: baryon acoustic oscillations from galaxies and quasars,''
JCAP \textbf{04} (2025), 012
[arXiv:2404.03000 [astro-ph.CO]].

\bibitem{DESI:2024lzq}
A.~G.~Adame \textit{et al.} [DESI],
``DESI 2024 IV: Baryon Acoustic Oscillations from the Lyman alpha forest,''
JCAP \textbf{01} (2025), 124
[arXiv:2404.03001 [astro-ph.CO]].

\bibitem{DESI:2024mwx}
A.~G.~Adame \textit{et al.} [DESI],
``DESI 2024 VI: cosmological constraints from the measurements of baryon acoustic oscillations,''
JCAP \textbf{02} (2025), 021
[arXiv:2404.03002 [astro-ph.CO]].

\bibitem{Chevallier:2000qy}
M.~Chevallier and D.~Polarski,
``Accelerating universes with scaling dark matter,''
Int. J. Mod. Phys. D \textbf{10} (2001), 213-224
[arXiv:gr-qc/0009008 [gr-qc]].

\bibitem{Linder:2002et}
E.~V.~Linder,
``Exploring the expansion history of the universe,''
Phys. Rev. Lett. \textbf{90} (2003), 091301
[arXiv:astro-ph/0208512 [astro-ph]].

\bibitem{Planck:2018vyg}
N.~Aghanim \textit{et al.} [Planck],
``Planck 2018 results. VI. Cosmological parameters,''
Astron. Astrophys. \textbf{641} (2020), A6
[erratum: Astron. Astrophys. \textbf{652} (2021), C4]
[arXiv:1807.06209 [astro-ph.CO]].

\bibitem{Brout:2022vxf}
D.~Brout, D.~Scolnic, B.~Popovic, A.~G.~Riess, J.~Zuntz, R.~Kessler, A.~Carr, T.~M.~Davis, S.~Hinton and D.~Jones, \textit{et al.}
``The Pantheon+ Analysis: Cosmological Constraints,''
Astrophys. J. \textbf{938} (2022) no.2, 110
[arXiv:2202.04077 [astro-ph.CO]].

\bibitem{Rubin:2023ovl}
D.~Rubin, G.~Aldering, M.~Betoule, A.~Fruchter, X.~Huang, A.~G.~Kim, C.~Lidman, E.~Linder, S.~Perlmutter and P.~Ruiz-Lapuente, \textit{et al.}
``Union Through UNITY: Cosmology with 2,000 SNe Using a Unified Bayesian Framework,''
[arXiv:2311.12098 [astro-ph.CO]].

\bibitem{DES:2024jxu}
T.~M.~C.~Abbott \textit{et al.} [DES],
``The Dark Energy Survey: Cosmology Results with {\ensuremath{\sim}}1500 New High-redshift Type Ia Supernovae Using the Full 5 yr Data Set,''
Astrophys. J. Lett. \textbf{973} (2024) no.1, L14
[arXiv:2401.02929 [astro-ph.CO]].

\bibitem{Foreman-Mackey:2012any}
D.~Foreman-Mackey, D.~W.~Hogg, D.~Lang and J.~Goodman,
``emcee: The MCMC Hammer,''
Publ. Astron. Soc. Pac. \textbf{125} (2013), 306-312
[arXiv:1202.3665 [astro-ph.IM]].

\bibitem{Colgain:2022nlb}
E.~\'O~Colg\'ain, M.~M.~Sheikh-Jabbari, R.~Solomon, G.~Bargiacchi, S.~Capozziello, M.~G.~Dainotti and D.~Stojkovic,
``Revealing intrinsic flat \ensuremath{\Lambda}CDM biases with standardizable candles,''
Phys. Rev. D \textbf{106} (2022) no.4, L041301
[arXiv:2203.10558 [astro-ph.CO]].

\bibitem{BOSS:2016wmc}
S.~Alam \textit{et al.} [BOSS],
``The clustering of galaxies in the completed SDSS-III Baryon Oscillation Spectroscopic Survey: cosmological analysis of the DR12 galaxy sample,''
Mon. Not. Roy. Astron. Soc. \textbf{470} (2017) no.3, 2617-2652
[arXiv:1607.03155 [astro-ph.CO]].


\bibitem{eBOSS:2020yzd}
S.~Alam \textit{et al.} [eBOSS],
``Completed SDSS-IV extended Baryon Oscillation Spectroscopic Survey: Cosmological implications from two decades of spectroscopic surveys at the Apache Point Observatory,''
Phys. Rev. D \textbf{103} (2021) no.8, 083533
[arXiv:2007.08991 [astro-ph.CO]].

\bibitem{DES:2024pwq}
T.~M.~C.~Abbott \textit{et al.} [DES],
``Dark Energy Survey: A 2.1\% measurement of the angular baryonic acoustic oscillation scale at redshift zeff=0.85 from the final dataset,''
Phys. Rev. D \textbf{110} (2024) no.6, 063515
[arXiv:2402.10696 [astro-ph.CO]].

\bibitem{Wong:2019kwg}
K.~C.~Wong, S.~H.~Suyu, G.~C.~F.~Chen, C.~E.~Rusu, M.~Millon, D.~Sluse, V.~Bonvin, C.~D.~Fassnacht, S.~Taubenberger and M.~W.~Auger, \textit{et al.}
``H0LiCOW -- XIII. A 2.4 per cent measurement of H0 from lensed quasars: 5.3${\sigma}$ tension between early- and late-Universe probes,''
Mon. Not. Roy. Astron. Soc. \textbf{498} (2020) no.1, 1420-1439
[arXiv:1907.04869 [astro-ph.CO]].

\bibitem{DES:2019fny}
A.~J.~Shajib \textit{et al.} [DES],
``STRIDES: a 3.9 per cent measurement of the Hubble constant from the strong lens system DES J0408--5354,''
Mon. Not. Roy. Astron. Soc. \textbf{494} (2020) no.4, 6072-6102
[arXiv:1910.06306 [astro-ph.CO]].

[arXiv:2009.06740]]
\bibitem{Millon:2019slk}
M.~Millon, A.~Galan, F.~Courbin, T.~Treu, S.~H.~Suyu, X.~Ding, S.~Birrer, G.~C.~F.~Chen, A.~J.~Shajib and D.~Sluse, \textit{et al.}
``TDCOSMO. I. An exploration of systematic uncertainties in the inference of $H_0$ from time-delay cosmography,''
Astron. Astrophys. \textbf{639} (2020), A101
[arXiv:1912.08027 [astro-ph.CO]].

\bibitem{Kelly:2023mgv}
P.~L.~Kelly, S.~Rodney, T.~Treu, M.~Oguri, W.~Chen, A.~Zitrin, S.~Birrer, V.~Bonvin, L.~Dessart and J.~M.~Diego, \textit{et al.}
``Constraints on the Hubble constant from supernova Refsdal's reappearance,''
Science \textbf{380} (2023) no.6649, abh1322
[arXiv:2305.06367 [astro-ph.CO]].

\bibitem{Pascale:2024qjr}
M.~Pascale, B.~L.~Frye, J.~D.~R.~Pierel, W.~Chen, P.~L.~Kelly, S.~H.~Cohen, R.~A.~Windhorst, A.~G.~Riess, P.~S.~Kamieneski and J.~M.~Diego, \textit{et al.}
``SN H0pe: The First Measurement of $H_0$ from a Multiply-Imaged Type Ia Supernova, Discovered by JWST,''
[arXiv:2403.18902 [astro-ph.CO]].

\bibitem{Li:2024elb}
X.~Li and K.~Liao,
``Determining Cosmological-model-independent $H_0$ with Gravitationally Lensed Supernova Refsdal,''
[arXiv:2401.12052 [astro-ph.CO]].

\bibitem{Dainotti:2021pqg}
M.~G.~Dainotti, B.~De Simone, T.~Schiavone, G.~Montani, E.~Rinaldi and G.~Lambiase,
``On the Hubble constant tension in the SNe Ia Pantheon sample,''
Astrophys. J. \textbf{912} (2021) no.2, 150
[arXiv:2103.02117 [astro-ph.CO]].

\bibitem{Colgain:2022rxy}
E.~\'O~Colg\'ain, M.~M.~Sheikh-Jabbari, R.~Solomon, M.~G.~Dainotti and D.~Stojkovic,
``Putting flat ${\Lambda}$CDM in the (Redshift) bin,''
Phys. Dark Univ. \textbf{44} (2024), 101464
[arXiv:2206.11447 [astro-ph.CO]].

\bibitem{Jia:2022ycc}
X.~D.~Jia, J.~P.~Hu and F.~Y.~Wang,
``Evidence of a decreasing trend for the Hubble constant,''
Astron. Astrophys. \textbf{674} (2023), A45
[arXiv:2212.00238 [astro-ph.CO]].

\bibitem{Pasten:2023rpc}
E.~Past\'en and V.~H.~C\'ardenas,
``Testing ${\Lambda}$CDM cosmology in a binned universe: Anomalies in the deceleration parameter,''
Phys. Dark Univ. \textbf{40} (2023), 101224
[arXiv:2301.10740 [astro-ph.CO]].

\bibitem{Malekjani:2023ple}
M.~Malekjani, R.~M.~Conville, E.~\'O~Colg\'ain, S.~Pourojaghi and M.~M.~Sheikh-Jabbari,
``On redshift evolution and negative dark energy density in Pantheon + Supernovae,''
Eur. Phys. J. C \textbf{84} (2024) no.3, 317
[arXiv:2301.12725 [astro-ph.CO]].


\bibitem{Wagner:2022etu}
J.~Wagner,
``Solving the Hubble tension \`a la Ellis \& Stoeger 1987,''
PoS \textbf{CORFU2022} (2023), 267
[arXiv:2203.11219 [astro-ph.CO]].

\bibitem{Hu:2022kes}
J.~P.~Hu and F.~Y.~Wang,
``Revealing the late-time transition of H$_0$: relieve the Hubble crisis,''
Mon. Not. Roy. Astron. Soc. \textbf{517} (2022) no.1, 576-581
[arXiv:2203.13037 [astro-ph.CO]].

\bibitem{Dainotti:2023yrk}
M.~Dainotti, B.~De Simone, G.~Montani;, T.~Schiavone; and G.~Lambiase.,
``The Hubble constant tension: current status and future perspectives through new cosmological probes,''
PoS \textbf{CORFU2022} (2023), 235
doi:10.22323/1.436.0235
[arXiv:2301.10572 [astro-ph.CO]].

\bibitem{Dainotti:2022bzg}
M.~G.~Dainotti, B.~De Simone, T.~Schiavone, G.~Montani, E.~Rinaldi, G.~Lambiase, M.~Bogdan and S.~Ugale,
``On the Evolution of the Hubble Constant with the SNe Ia Pantheon Sample and Baryon Acoustic Oscillations: A Feasibility Study for GRB-Cosmology in 2030,''
Galaxies \textbf{10} (2022) no.1, 24
[arXiv:2201.09848 [astro-ph.CO]].

\bibitem{Krishnan:2020obg}
C.~Krishnan, E.~\'O~Colg\'ain, Ruchika, A.~A.~Sen, M.~M.~Sheikh-Jabbari and T.~Yang,
``Is there an early Universe solution to Hubble tension?,''
Phys. Rev. D \textbf{102} (2020) no.10, 103525
[arXiv:2002.06044 [astro-ph.CO]].

\bibitem{Dainotti:2022rea}
M.~G.~Dainotti, G.~Sarracino and S.~Capozziello,
``Gamma-ray bursts, supernovae Ia, and baryon acoustic oscillations: A binned cosmological analysis,''
Publ. Astron. Soc. Jap. \textbf{74} (2022) no.5, 1095-1113-1113
[arXiv:2206.07479 [astro-ph.CO]].

\bibitem{Risaliti:2018reu}
G.~Risaliti and E.~Lusso,
``Cosmological constraints from the Hubble diagram of quasars at high redshifts,''
Nature Astron. \textbf{3} (2019) no.3, 272-277
[arXiv:1811.02590 [astro-ph.CO]].

\bibitem{Lusso:2020pdb}
E.~Lusso, G.~Risaliti, E.~Nardini, G.~Bargiacchi, M.~Benetti, S.~Bisogni, S.~Capozziello, F.~Civano, L.~Eggleston and M.~Elvis, \textit{et al.}
``Quasars as standard candles III. Validation of a new sample for cosmological studies,''
Astron. Astrophys. \textbf{642} (2020), A150
[arXiv:2008.08586 [astro-ph.GA]].

\bibitem{Dainotti:2008vw}
M.~G.~Dainotti, V.~F.~Cardone and S.~Capozziello,
``A time - luminosity correlation for Gamma Ray Bursts in the X - rays,''
Mon. Not. Roy. Astron. Soc. \textbf{391} (2008), 79
[arXiv:0809.1389 [astro-ph]].

\bibitem{Dainotti:2022rfz}
M.~G.~Dainotti, G.~Bargiacchi, A.~L.~Lenart, S.~Capozziello, E.~\'O~Colg\'ain, R.~Solomon, D.~Stojkovic and M.~M.~Sheikh-Jabbari,
``Quasar Standardization: Overcoming Selection Biases and Redshift Evolution,''
Astrophys. J. \textbf{931} (2022) no.2, 106
[arXiv:2203.12914 [astro-ph.HE]].

\bibitem{Dainotti:2023cpn}
M.~G.~Dainotti, G.~Bargiacchi, A.~\L{}.~Lenart, S.~Nagataki and S.~Capozziello,
``Quasars: Standard Candles up to z = 7.5 with the Precision of Supernovae Ia,''
Astrophys. J. \textbf{950} (2023) no.1, 45
[arXiv:2305.19668 [astro-ph.CO]].

\bibitem{Pourojaghi:2022zrh}
S.~Pourojaghi, N.~F.~Zabihi and M.~Malekjani,
``Can high-redshift Hubble diagrams rule out the standard model of cosmology in the context of cosmography?,''
Phys. Rev. D \textbf{106} (2022) no.12, 123523
[arXiv:2212.04118 [astro-ph.CO]].

\bibitem{Bargiacchi:2023jse}
G.~Bargiacchi, M.~G.~Dainotti, S.~Nagataki and S.~Capozziello,
``Gamma-ray bursts, quasars, baryonic acoustic oscillations, and supernovae Ia: new statistical insights and cosmological constraints,''
Mon. Not. Roy. Astron. Soc. \textbf{521} (2023) no.3, 3909-3924
[arXiv:2303.07076 [astro-ph.CO]].

\bibitem{Vagnozzi:2023nrq}
S.~Vagnozzi,
``Seven Hints That Early-Time New Physics Alone Is Not Sufficient to Solve the Hubble Tension,''
Universe \textbf{9} (2023) no.9, 393
[arXiv:2308.16628 [astro-ph.CO]].

\bibitem{Khadka:2020vlh}
N.~Khadka and B.~Ratra,
``Using quasar X-ray and UV flux measurements to constrain cosmological model parameters,''
Mon. Not. Roy. Astron. Soc. \textbf{497} (2020) no.1, 263-278
[arXiv:2004.09979 [astro-ph.CO]].

\bibitem{Khadka:2020tlm}
N.~Khadka and B.~Ratra,
``Determining the range of validity of quasar X-ray and UV flux measurements for constraining cosmological model parameters,''
Mon. Not. Roy. Astron. Soc. \textbf{502} (2021) no.4, 6140-6156
[arXiv:2012.09291 [astro-ph.CO]].

\bibitem{Singal:2022nto}
J.~Singal, S.~Mutchnick and V.~Petrosian,
``The X-Ray Luminosity Function Evolution of Quasars and the Correlation between the X-Ray and Ultraviolet Luminosities,''
Astrophys. J. \textbf{932} (2022) no.2, 111
[arXiv:2203.13374 [astro-ph.CO]].

\bibitem{Petrosian:2022tlp}
V.~Petrosian, J.~Singal and S.~Mutchnick,
``Can the Distance-Redshift Relation be Determined from Correlations between Luminosities?,''
Astrophys. J. Lett. \textbf{935} (2022) no.1, L19
[arXiv:2205.07981 [astro-ph.CO]].

\bibitem{Zajacek:2023qjm}
M.~Zaja\v{c}ek, B.~Czerny, N.~Khadka, M.~L.~Mart\'\i{}nez-Aldama, R.~Prince, S.~Panda and B.~Ratra,
``Effect of extinction on quasar luminosity distances determined from UV and X-ray flux measurements,''
[arXiv:2305.08179 [astro-ph.GA]].

\bibitem{Dainotti:2024aha}
M.~G.~Dainotti, A.~L.~Lenart, M.~G.~Yengejeh, S.~Chakraborty, N.~Fraija, E.~Di Valentino and G.~Montani,
``A new binning method to choose a standard set of Quasars,''
Phys. Dark Univ. \textbf{44} (2024), 101428
[arXiv:2401.12847 [astro-ph.HE]].

\bibitem{Dainotti:2024bth}
M.~G.~Dainotti, G.~Bargiacchi, A.~\L{}.~Lenart and S.~Capozziello,
``The Scavenger Hunt for Quasar Samples to Be Used as Cosmological Tools,''
Galaxies \textbf{12} (2024) no.1, 4
[arXiv:2401.11998 [astro-ph.CO]].

\bibitem{Lenart:2022nip}
A.~\L{}.~Lenart, G.~Bargiacchi, M.~G.~Dainotti, S.~Nagataki and S.~Capozziello,
``A Bias-free Cosmological Analysis with Quasars Alleviating H $_{0}$ Tension,''
Astrophys. J. Suppl. \textbf{264} (2023) no.2, 46
[arXiv:2211.10785 [astro-ph.CO]].

\bibitem{Colgain:2024ksa}
E.~{\'O}~Colg{\'a}in, S.~Pourojaghi and M.~M.~Sheikh-Jabbari,
``Implications of DES 5YR SNe Dataset for $\Lambda $CDM,''
Eur. Phys. J. C \textbf{85} (2025) no.3, 286
[arXiv:2406.06389 [astro-ph.CO]].

\bibitem{Demianski:2016zxi}
M.~Demianski, E.~Piedipalumbo, D.~Sawant and L.~Amati,
``Cosmology with gamma-ray bursts: I. The Hubble diagram through the calibrated $E_{\rm p,i}$ - $E_{\rm iso}$ correlation,''
Astron. Astrophys. \textbf{598} (2017), A112
[arXiv:1610.00854 [astro-ph.CO]].

\bibitem{Demianski:2016dsa}
M.~Demianski, E.~Piedipalumbo, D.~Sawant and L.~Amati,
Astron. Astrophys. \textbf{598} (2017), A113
[arXiv:1609.09631 [astro-ph.CO]].

\bibitem{Dainotti:2022ked}
M.~G.~Dainotti, A.~\L{}.~Lenart, A.~Chraya, G.~Sarracino, S.~Nagataki, N.~Fraija, S.~Capozziello and M.~Bogdan,
``The Gamma-ray Bursts fundamental plane correlation as a cosmological tool,''
[arXiv:2209.08675 [astro-ph.HE]].

\bibitem{Khadka:2021vqa}
N.~Khadka, O.~Luongo, M.~Muccino and B.~Ratra,
``Do gamma-ray burst measurements provide a useful test of cosmological models?,''
JCAP \textbf{09} (2021), 042
[arXiv:2105.12692 [astro-ph.CO]].

\bibitem{Alfano:2024ukk}
A.~C.~Alfano, S.~Capozziello, O.~Luongo and M.~Muccino,
``Cosmological transition epoch from gamma-ray burst correlations,''
[arXiv:2402.18967 [astro-ph.CO]].

\bibitem{Srinivasaragavan:2020isz}
G.~P.~Srinivasaragavan, M.~G.~Dainotti, N.~Fraija, X.~Hernandez, S.~Nagataki, A.~Lenart, L.~Bowden and R.~Wagner,
``On the investigation of the closure relations for Gamma-Ray Bursts observed by Swift in the post-plateau phase and the GRB fundamental plane,''
Astrophys. J. \textbf{903} (2020) no.1, 18
[arXiv:2009.06740 [astro-ph.HE]].

\bibitem{Colgain:2024mtg}
E.~{\'O}~Colg{\'a}in and M.~M.~Sheikh-Jabbari,
``DESI and SNe: Dynamical Dark Energy, $\Omega_m$ Tension or Systematics?,''
[arXiv:2412.12905 [astro-ph.CO]].

\bibitem{DESI:2016fyo}
A.~Aghamousa \textit{et al.} [DESI],
``The DESI Experiment Part I: Science, Targeting, and Survey Design,''
[arXiv:1611.00036 [astro-ph.IM]].

\bibitem{Alcock:1979mp}
C.~Alcock and B.~Paczynski,
``An evolution free test for non-zero cosmological constant,''
Nature \textbf{281} (1979), 358-359



\bibitem{Dainotti:2023bwq}
M.~G.~Dainotti, G.~Bargiacchi, M.~Bogdan, A.~\L{}.~Lenart, K.~Iwasaki, S.~Capozziello, B.~Zhang and N.~Fraija,
``Reducing the Uncertainty on the Hubble Constant up to 35\% with an Improved Statistical Analysis: Different Best-fit Likelihoods for Type Ia Supernovae, Baryon Acoustic Oscillations, Quasars, and Gamma-Ray Bursts,''
Astrophys. J. \textbf{951} (2023) no.1, 63
[arXiv:2305.10030 [astro-ph.CO]].

\bibitem{Lewis:2019xzd}
A.~Lewis,
``GetDist: a Python package for analysing Monte Carlo samples,''
[arXiv:1910.13970 [astro-ph.IM]].

\bibitem{Cortes:2024lgw}
M.~Cort\^es and A.~R.~Liddle,
``Interpreting DESI's evidence for evolving dark energy,''
[arXiv:2404.08056 [astro-ph.CO]].

\bibitem{Wang:2024rjd}
D.~Wang,
``The Self-Consistency of DESI Analysis and Comment on ''Does DESI 2024 Confirm $\Lambda$CDM?'',''
[arXiv:2404.13833 [astro-ph.CO]].

\bibitem{Wang:2024pui}
Z.~Wang, S.~Lin, Z.~Ding and B.~Hu,
``The role of LRG1 and LRG2's monopole in inferring the DESI 2024 BAO cosmology,''
[arXiv:2405.02168 [astro-ph.CO]].

\bibitem{Chudaykin:2024gol}
A.~Chudaykin and M.~Kunz,
``Modified gravity interpretation of the evolving dark energy in light of DESI data,''
[arXiv:2407.02558 [astro-ph.CO]].


\bibitem{Tada:2024znt}
Y.~Tada and T.~Terada,
``Quintessential interpretation of the evolving dark energy in light of DESI,''
[arXiv:2404.05722 [astro-ph.CO]].

\bibitem{Gu:2024jhl}
G.~Gu, X.~Wang, X.~Mu, S.~Yuan and G.~B.~Zhao,
``Dynamical dark energy in light of cosmic distance measurements I: a demonstration using simulated datasets,''
[arXiv:2404.06303 [astro-ph.CO]].

\bibitem{Wang:2024qan}
X.~Wang, G.~Gu, X.~Mu, S.~Yuan and G.~B.~Zhao,
``Dynamical dark energy in light of cosmic distance measurements II: a study using current observations,''
[arXiv:2404.06310 [astro-ph.CO]].

\bibitem{Wang:2024hks}
D.~Wang,
``Constraining Cosmological Physics with DESI BAO Observations,''
[arXiv:2404.06796 [astro-ph.CO]].

\bibitem{Luongo:2024fww}
O.~Luongo and M.~Muccino,
``Model independent cosmographic constraints from DESI 2024,''
[arXiv:2404.07070 [astro-ph.CO]].

\bibitem{Yin:2024hba}
W.~Yin,
``Cosmic Clues: DESI, Dark Energy, and the Cosmological Constant Problem,''
[arXiv:2404.06444 [hep-ph]].

\bibitem{Colgain:2019pck}
E.~\'O~Colg\'ain,
``A hint of matter underdensity at low $z$?,''
JCAP \textbf{09} (2019), 006
[arXiv:1903.11743 [astro-ph.CO]].

\bibitem{Kazantzidis:2020xta}
L.~Kazantzidis, H.~Koo, S.~Nesseris, L.~Perivolaropoulos and A.~Shafieloo,
``Hints for possible low redshift oscillation around the best-fitting $\Lambda$CDM model in the expansion history of the Universe,''
Mon. Not. Roy. Astron. Soc. \textbf{501} (2021) no.3, 3421-3426
[arXiv:2010.03491 [astro-ph.CO]].

\bibitem{Amon:2022azi}
A.~Amon and G.~Efstathiou,
``A non-linear solution to the $S_8$ tension?,''
doi:10.1093/mnras/stac2429
[arXiv:2206.11794 [astro-ph.CO]].

\bibitem{Preston:2023uup}
C.~Preston, A.~Amon and G.~Efstathiou,
``A non-linear solution to the S8 tension-- II. Analysis of DES Year 3 cosmic shear,''
Mon. Not. Roy. Astron. Soc. \textbf{525} (2023) no.4, 5554-5564
[arXiv:2305.09827 [astro-ph.CO]].

\bibitem{Terasawa:2024agq}
R.~Terasawa, X.~Li, M.~Takada, T.~Nishimichi, S.~Tanaka, S.~Sugiyama, T.~Kurita, T.~Zhang, M.~Shirasaki and R.~Takahashi, \textit{et al.}
``Exploring the baryonic effect signature in the Hyper Suprime-Cam Year 3 cosmic shear two-point correlations on small scales: the $S_8$ tension remains present,''
[arXiv:2403.20323 [astro-ph.CO]].

\bibitem{Esposito:2022plo}
M.~Esposito, V.~Ir\v{s}i\v{c}, M.~Costanzi, S.~Borgani, A.~Saro and M.~Viel,
``Weighing cosmic structures with clusters of galaxies and the intergalactic medium,''
Mon. Not. Roy. Astron. Soc. \textbf{515} (2022) no.1, 857-870
[arXiv:2202.00974 [astro-ph.CO]].

\bibitem{Adil:2023jtu}
S.~A.~Adil, \"O.~Akarsu, M.~Malekjani, E.~\'O~Colg\'ain, S.~Pourojaghi, A.~A.~Sen and M.~M.~Sheikh-Jabbari,
``$S_8$ increases with effective redshift in ${\Lambda}$CDM cosmology,''
Mon. Not. Roy. Astron. Soc. \textbf{528} (2023) no.1, L20-L26
[arXiv:2303.06928 [astro-ph.CO]].

\bibitem{ACT:2023kun}
M.~S.~Madhavacheril \textit{et al.} [ACT],
``The Atacama Cosmology Telescope: DR6 Gravitational Lensing Map and Cosmological Parameters,''
Astrophys. J. \textbf{962} (2024) no.2, 113
[arXiv:2304.05203 [astro-ph.CO]].

\bibitem{Tutusaus:2023aux}
I.~Tutusaus, C.~Bonvin and N.~Grimm,
``First measurement of the Weyl potential evolution from the Year 3 Dark Energy Survey data: Localising the $\sigma_8$ tension,''
[arXiv:2312.06434 [astro-ph.CO]].

\bibitem{DESI:2024hhd}
A.~G.~Adame \textit{et al.} [DESI],
``DESI 2024 VII: Cosmological Constraints from the Full-Shape Modeling of Clustering Measurements,''
[arXiv:2411.12022 [astro-ph.CO]].

\bibitem{DESI:2024jis}
A.~G.~Adame \textit{et al.} [DESI],
``DESI 2024 V: Full-Shape Galaxy Clustering from Galaxies and Quasars,''
[arXiv:2411.12021 [astro-ph.CO]].

\bibitem{DESI:2025zgx}
M.~Abdul Karim \textit{et al.} [DESI],
``DESI DR2 Results II: Measurements of Baryon Acoustic Oscillations and Cosmological Constraints,''
[arXiv:2503.14738 [astro-ph.CO]].

\bibitem{Colgain:2025nzf}
E.~{\'O}~Colg{\'a}in, S.~Pourojaghi, M.~M.~Sheikh-Jabbari and L.~Yin,
``How much has DESI dark energy evolved since DR1?,''
[arXiv:2504.04417 [astro-ph.CO]].

\end{thebibliography}

\end{document}